\documentclass[12pt]{article}
\usepackage{graphicx}

\usepackage{relsize}
\usepackage{xspace}
\def\babar{\mbox{\slshape B\kern-0.1em{\smaller A}\kern-0.1em
    B\kern-0.1em{\smaller A\kern-0.2em R}}\xspace}

\newcommand\pubnumber{}
\newcommand\pubdate{\today}

\def\queenmary{Queen Mary, University of London,\\
Department of Physics, Mile End Road,\\
London, E1 4NS, UK}
\def\support{\footnote{on behalf of the Super$B$ Collaboration.}}

\def\Title#1{\begin{center} {\Large #1 } \end{center}}
\def\Author#1{\begin{center}{ \sc #1} \end{center}}
\def\Address#1{\begin{center}{ \it #1} \end{center}}

\newcommand\pubblock{\rightline{\begin{tabular}{l} \pubnumber\\
         \pubdate  \end{tabular}}}

\def\beq{\begin{equation}}
\def\eeq#1{\label{#1}\end{equation}}
\def\eeqn{\end{equation}}

\def\beqa{\begin{eqnarray}}
\def\eeqa#1{\label{#1}\end{eqnarray}}
\def\eeqan{\end{eqnarray}}

\let\bar=\overbar

\def\Dslash{\not{\hbox{\kern-4pt $D$}}}
\def\dslash{\not{\hbox{\kern-2pt $\del$}}}

\def\msb{{\bar{\ssstyle M \kern -1pt S}}}

\begin{document}
\begin{titlepage}
\pubblock

\vfill
\Title{Charm Mesons at the Super$B$ Experiment: Rare Decays, Mixing and {$CP$} Violation Potential}
\vfill
\Author{ Gianluca Inguglia \support}
\Address{\queenmary}
\vfill
\begin{abstract}
%
%
The Super$B$ experiment at the Cabibbo Laboratory will provide new possibilities to study the physics of $charm$. The potential physics reach of the experiment when performing studies of rare decays, mixing and $CP$ violation in charm decays is presented here and the implications of such measurements for new physics scenarios is discussed. 

\end{abstract}



\vfill


\end{titlepage}

\setcounter{footnote}{0}

\section{Introduction}
\label{sec:intro}

Charm mesons are bound states of a quark-antiquark pair in which one, the quark or the antiquark, is a charm ($c$) quark. The charm quark was remarkably discovered in 1974 when two different teams and experiments, Stanford Linear Accelerator Center (SLAC) and Brookhaven National Laboratory (BNL) announced the existence of a new resonance with a mass of almost $3.1$ $GeV$~\cite{SLAC}\cite{BNL}. Since the mass of this state was to large to be explained in terms of the three known light quarks ($u, d, s$), the existence of a new heavy quark was needed to physically interpret the new particle. The announcement of the discovery was made simultaneously by the two teams in November 1974, and it is often referred to as the $November$ $Revolution$. However, before this discovery, the GIM mechanism was proposed, in which the absence of flavour changing neutral currents (FCNCs) was considered as an indication of the existence of a fourth quark to explain the suppression of some transitions~\cite{GIM}.
FCNCs still represent a very important issue in flavour physics when looking for new physics. Another interesting aspect is that the standard model (SM) predicts small mixing and $CP$ violation for charm mesons. While mixing has already been observed for $D^0$ mesons, $CP$ violation in charm has still to be discovered and a time-dependent analysis may provide a tool to test the Cabibbo-Kobayashi-Maskawa mechanism~\cite{cabibbo}\cite{kob} and the SM itself through the measurement of the angle $\beta_{c,eff}$ in the charm unitarity triangle. 
This work describes the potential reach of Super$B$ when studying rare decays, mixing and $CP$ violation in charm mesons.
It is useful to stress here on the uniqueness of charm mesons in the understanding of the SM. $D^0$ mesons are the only mesons made from two $up-type$ quarks ($c$ and $\bar u$), this allows one to better understand the flavour changing structure of the SM, and moreover it may prove (or not) the $CP$ violation interpretation in terms of the CKM matrix in the $up$-sector (a missing piece of information in our understanding of flavour physics). 
\section{Rare Decays}
As discussed in the introduction, the GIM mechanism requires the existence of a fourth quark, the charm quark. Due to GIM suppression, the SM predicts very low rates for FCNC decays, so that the study of rare charm decays becomes interesting not only for understanding the charm quark, but also as a tool to test the SM and to look for new physics (NP). New undiscovered heavy particles may play a role for example in the loop diagrams involved in rare $D$ decays, this may allow one to indirectly prove the presence of such particles by studying the deviations from the predicted SM rate for a particular rare $D^0$ decay mode. This approach does not require a very high energy facility, where the higher energy available may allow one to directly produce the new particles, but needs a high luminosity facility. There are several rare decays that can be studied (see for example~\cite{cdr}), here the decays $D^0 \to \gamma\gamma$, $D^0 \to \mu^+\mu^-$ and $c \to u l^+l^-$ are described.

\subsection{$D^0 \to \gamma\gamma$}
In the SM the process $D^0 \to \gamma\gamma$ represents a FCNC which is forbidden at tree level, and which is dominated by long distance (LD) contributions. Short distance (SD) and LD contributions can be calculated, and for a mass of the charm quark satisfying $M_D-m_c \approx 300 MeV$ one obtains~\cite{vmd}:
\begin{eqnarray}
Br_{D^0 \to \gamma\gamma}^{SD} \approx 3 \times 10^{-11},  \label{EQ:sdgammagmamma}
\end{eqnarray}
\begin{eqnarray}
Br_{D^0 \to \gamma\gamma}^{LD} \approx 1 \pm 0.5 \times 10^{-8}.  \label{EQ:ldgammagmamma}
\end{eqnarray}

\begin{figure}[!ht]
\begin{center}
\resizebox{10.cm}{!}{
\includegraphics{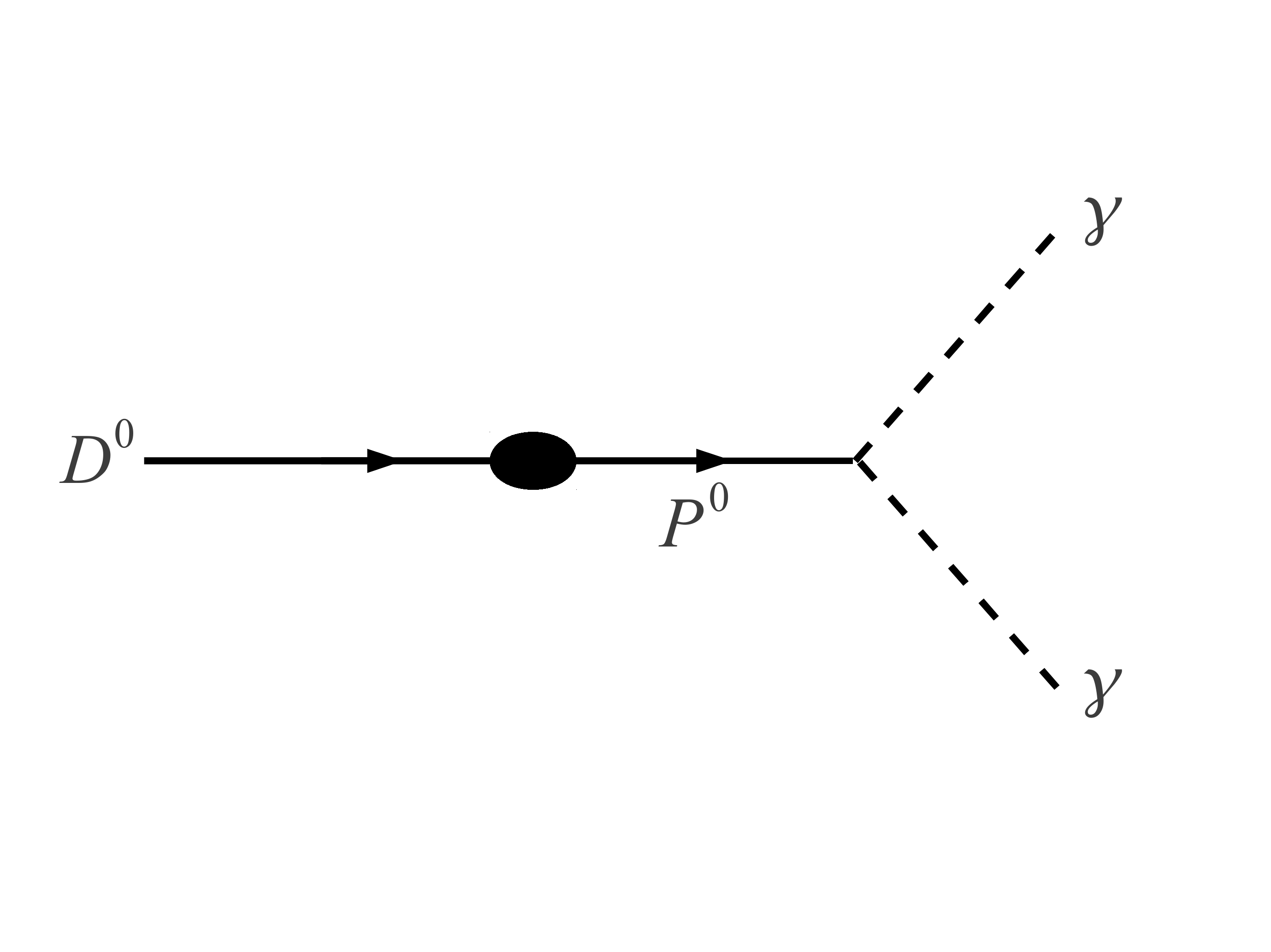}
}
\caption{The vector meson dominance contribution for $D^0 \to \gamma \gamma$.}\label{fig:vector}
\end{center}
\end{figure}

LD contributions need to be modelled in order to estimate the amplitude $D^0 \to \gamma\gamma$. It is possible to parametrise the decay as if it was driven by a single vector meson dominance process in which the $D^0$ weakly mixes with a vector meson $P^0$ as shown in Fig.~\ref{fig:vector}~\cite{vmd}.

Limits on this decay channel have been obtained by the CLEO-c and \babar Collaborations: $B(D^0 \to \gamma\gamma)<2.7 \times 10^{-5}$ (CLEO-c)~\cite{CLEO} and $B(D^0 \to \gamma\gamma)<2.2 \times 10^{-6}$ (\babar)~\cite{BABAR}. The BESIII experiment is collecting data at the $\Psi(3770)$ and could reach a limit of $0.5 \times 10^{-7}$ with $20$ $fb^{-1}$ of data~\cite{BES}. The Super$B$ experiment is expected to collect $1.0$ $ab^{-1}$ at the charm threshold which translates into an  expected limit of the order of $1.0 \times 10^{-8}$ for $D^0 \to \gamma\gamma$, however studies are ongoing. 
\subsection{$D^0 \to \mu^+\mu^-$}
The rare decay $D^0 \to \mu^+\mu^-$ is a very important mode to be used when studying $\Delta C=1$ weak neutral currents. As for the decay $D^0 \to \gamma\gamma$, understanding of short and long distance contributions to this decay mode is necessary. Theoretical limits on this decay have been evaluated in Ref.~\cite{burdman}, as:
\begin{eqnarray}
Br(D^0 \to \mu^+\mu^-) \approx BR(D^0 \to \mu^+\mu^-)_{LD} &=& \nonumber \\ =3\times 10^{-5}Br(D^0 \to \gamma\gamma).  \label{EQ:mumu}
\end{eqnarray}
SD contributions for this decay are expected to be of the order of $10^{-18}$. NP enhancements may be manifest, but the LD contribution need to be understood and $under$ $control$ when interpreting any observed signal. The BESIII Collaboration may obtain a precision of $0.5 \times 10^{-7}$ with $20 fb^{-1}$ of data collected at the charm threshold. LHCb has already obtained the limit $Br(D^0 \to \mu^+\mu^-)<1.3 \times 10^{-8}$ with $0.9$ $fb^{-1}$ of data collected in 2011 at a center-of-mass energy of 7 TeV~\cite{lhcb}. Super$B$ expects to reach a precision of the order of $1.0 \times 10^{-8}$ when considering only one month of data collected at the charm threshold. 
\subsection{$c \to u l^+l^-$}
The decays $c \to u l^+l^-$ represent another class of interesting modes when performing tests for $\Delta C =1$ weak neutral currents, and the decays of interest are $D^0 \to \rho^0 l^+ l^-$ and $D^0 \to \pi^0 l^+ l^-$ and $D^0 \to \pi^+ l^+ l^-$, $D^0 \to \pi^0 e^\pm \mu^\mp $, $D^0 \to h^- l^+ l^+$ and $D^+ \to h^- e^\pm \mu^\mp$. Preliminary studies for these modes are ongoing for the Super$B$ Collaboration and the expected sensitivities for some of these decay modes are given in Table~\ref{tbl:ull}.
\begin{table}[!ht]
\caption{Expected sensitivities at Super$B$ and relative branching ratios for $D^0 \to \pi^0 l^+ l^-$ and $D^0 \to \pi^+ l^+ l^-$.}
\label{tbl:ull}
\renewcommand{\arraystretch}{1.2}
\begin{center}
    \begin{tabular}{ | l | c  c  |}
    \hline
    Decay mode 	&      	Sensitivity		&	BR (theory)       \\ \hline \hline
    $D^0 \to \pi^0 l^+ l^-$	&	$2 \times 10^{-8}$	&	$0.8 \times 10^{-6}$ \\
    $D^0 \to \pi^+ l^+ l^-$	&	$2 \times 10^{-8}$	&	$0.8 \times 10^{-6}$ \\
		
    \hline
    \end{tabular}
\end{center}
\end{table}

In Ref.~\cite{burdman} it has also been shown that the decay channels $D^0 \to \rho^0 e^+ e^-$ and $D^0 \to \rho^0 e^+ e^-$ may be used to evaluate the $strength$ of the contribution due to short and long distance effects and to constrain some supersymmetric (SUSY) models. In conclusion the transitions $c \to u l^+l^-$ are very useful also when looking for NP.

\section{Mixing}
\begin{figure}[!ht]
\begin{center}
\resizebox{10.cm}{!}{
\includegraphics{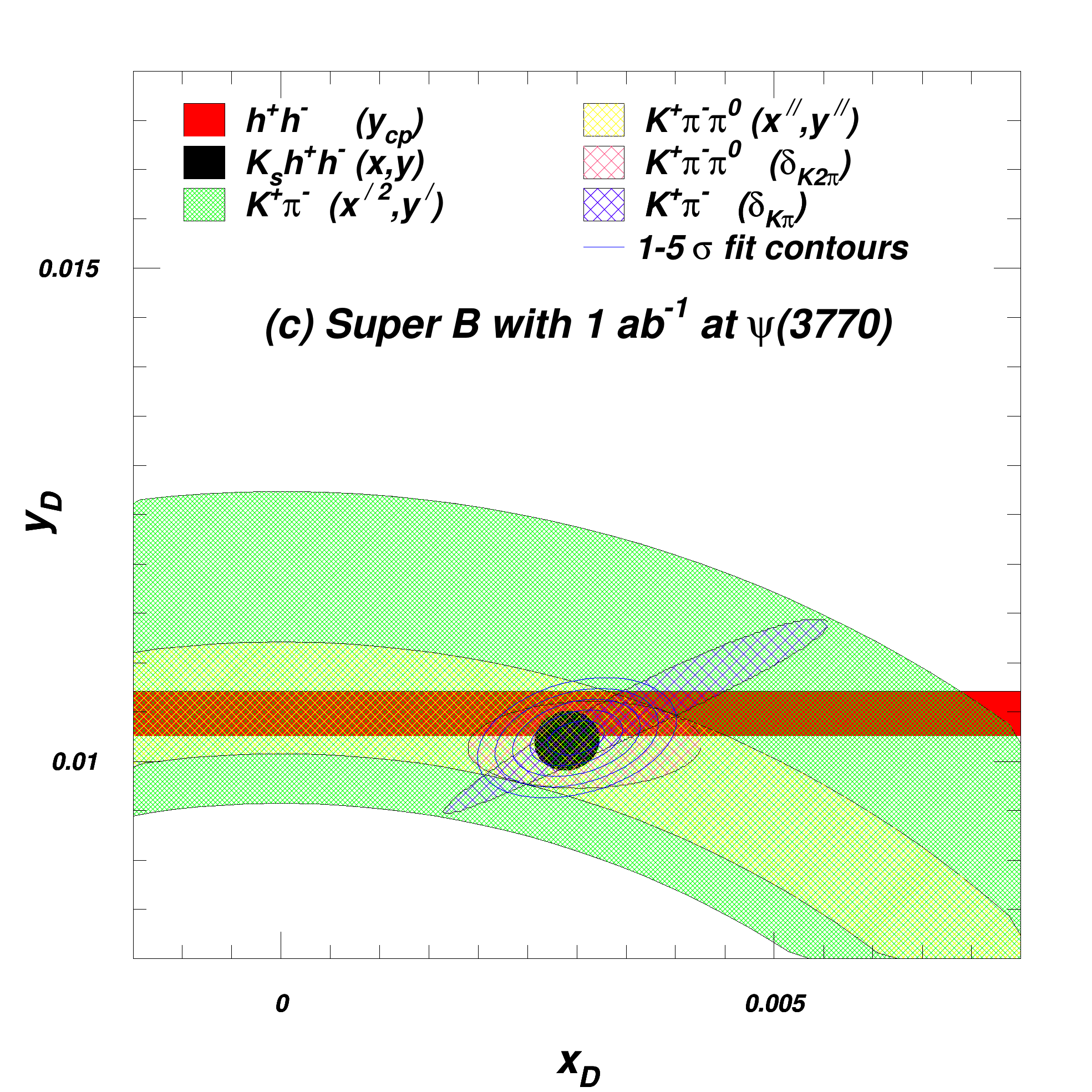}
}
\caption{Main charm mixing parameters combined into average values for $x_D$ and $y_D$ when considering a $1.0 ab^{-1}$ of data collected at the charm threshold (including projections of strong phase measurements $\delta_{K\pi}$ and $\delta_{k\pi\pi}$) and $75 ab^{-1}$ of data collected at the $\Upsilon(4S)$. The contours range from 1 to 4 standard deviations two-dimensional confidence regions from the $\chi^2$ fit to these results are shown as solid lines.}\label{fig:mixing2}
\end{center}
\end{figure}

Charm mixing was established by \babar and Belle through comparison of wrong sign (WS) decays $D^0 \to K^+\pi^-$ which are doubly Cabibbo suppressed (DCS) and the right sign (RS) Cabibbo favored (CF) decay $D^0 \to K^- \pi^+$~\cite{mixingbabar}\cite{mixingbelle}. However the precision with which the parameters $x_D$ and $y_D$ have been measured can still be improved. A better precision in the determination of the mixing parameters may allow an improved understanding of mixing in the $up$-sector and, if the $CP$ symmetry is broken in mixing. In fact the uncertainties on $x_D$ and $y_D$ are of the order of $2 \times 10^{-3}$, which is still to large to evaluate if there is any $CP$ difference between $D^0$ and $\bar D^0$.
Different measurements of charm mixing may be combined and projected into ($x_D,y_D$)~\cite{superb1}, as shown in Fig.~\ref{fig:mixing2}. This would require:
\begin{itemize}
  \item $\chi^2$ minimization technique
  \item Correlation effects need to be considered
  \begin{itemize}
          \item[-] ($x'^{2},y'$) from WS $D^0 \to K^+\pi^-$ decays 
          \item[-] ($x'',y''$) from time-dependent Dalitz plot (TDDP) analysis of $D^0 \to K^+ \pi^- \pi^0$
          \item[-] $y_{CP}$ from tagged/untagged $D^0\to h^+h^-$
          \item[-] ($x_D,y_D$) from the combined golden channels
        \end{itemize}
  \item $CP$ conserving hypothesis
\end{itemize}
The results of this analysis are given in Table~\ref{tbl:param}.
\begin{table}[!ht]
\begin{center}
\caption{Mixing parameters ($x_D$,$y_D$) and strong phases $\delta_{K\pi}$ and $\delta_{k\pi\pi}$ obtained from $\chi^2$ fit to observables obtained from Super$B$ when a 1.0 $ab^{-1}$ of data collected at the charm threshold is considered. The central value is arbitrarily chosen, and for this reason it is not shown.}\label{tbl:param}
\begin{tabular}{lcccc}
\\ \hline
Fit   & 
$ x \times 10^{-3}$                  & 
$ y \times 10^{-3}$                  & 
$\delta_{K^+\pi^-}^{\circ}$          & 
$\delta_{K^+\pi^-\pi^0}^{\circ}$     \\  \hline \\
  & $  xxx^{+  0.19}_{  -0.19}
$ & $ yyy^{+  0.11}_{  -0.11}
$ & $ \delta\delta\delta^{+  0.71}_{  -0.71}
$ & $ \delta\delta\delta^{+  0.83}_{  -0.83}
$ \\ \hline
\end{tabular}
\end{center}
\end{table}

It is clear that the Super$B$ experiment not only will improve our knowledge on charm mixing using a large data sample collected at the $\Upsilon(4S)$, but with the machine running at the charm threshold it will be possible to increase the sensitivity to the mixing parameters. In fact, with a run at charm threshold, one can reduces the
size of the $D^0 \to K_s^0h^+h^-$ ellipse. This reduces the area of the WS $D^0 \to K^+\pi^-$ ellipse that combines
strong phase measurement and $\Upsilon(4S)$ analysis of time-distribution of $K^+\pi^-$ decays, and reduces the area of the WS $D^0 \to K^+\pi^-\pi^0$ decays. Another interesting aspect of these studies is the possibility to define the $charm$ $golden$ $channels$ $D^0 \to K_S^0h^+h^- (h=\pi,K)$ which are self-conjugated multi-body final states and represent a combination of $CP$ odd and even eigenstates. If the measurement of strong phases yields $\delta_f=0,\pi$ then the parameters $x_D$ and $y_D$ are directly measurable with a TDDP analysis. 
\section{\boldmath{$CP$} Violation}
$CP$ violation was first observed in the Kaon system~\cite{Fitch} and in the $B$ meson system~\cite{CPB}\cite{CPBe}. In the charm sector, as anticipated previously, $CP$ violation has not been discovered yet and is expected to be small. 
$CP$ violation can be classified in three types: $direct$ $CP$ violation (in the decay), $indirect$ $CP$ violation (in mixing), and $CP$ violation in the $interference$ between mixing and decay (indirect, time-dependent).  
Recently the LHCb collaboration has reported a difference in (predominantly) direct $CP$ asymmetries in $D^0\to K^+ K^-$ and $D^0\to \pi^+ \pi^-$ that is $3.5 \sigma$ from the $CP$ conserving hypothesis~\cite{note}. This latest result makes $CP$ studies in charm very intriguing.
\subsection{Indirect \boldmath{$CP$} Violation}
Indirect $CP$ violation may be manifest through asymmetries in a broad class of observables. Here the approach which uses $effective$ $values$ of the mixing parameters is discussed~\cite{superb1}. 
Effective values of the mixing parameters are defined by considering separately the $D^0$ and the $\bar D^0$ mesons. One would then have 
\begin{eqnarray}
D^0 \to (x_D^+,y_D^+), \\
\bar D^0 \to (x_D^-,y_D^-),  \label{EQ:mumu}
\end{eqnarray}
where the sign $\pm$ depends upon the electric charge of the charm quark ($Q_c=+2/3$ and $Q_{\bar c}=-2/3$).
Ignoring systematic uncertainties (which will be almost identical for both the $D^0$'s and can then be neglected), if there is a difference $x_D^+-x_D^-$ of the order of $5 \times 10^{-4}$ (or larger) in the average $x$ value (or a difference $y_D^+-y_D^-$ of the order of $3 \times 10^{-4}$ (or larger) in the average $y$ value) Super$B$ will be able to measure it at a $3\sigma$ level. If these differences will be observed and interpreted as being due to $CP$ violation in mixing, then they would provide a measurement of $|\frac{q_D}{p_D}|$.
If one neglects direct $CP$ violation, then $x_D^+ \approx |\frac{q_D}{p_D}|x_D$ and $x_D^- \approx |\frac{p_D}{q_D}|x_D$, and very similar relations may be found for $y_D^+$ and $y_D^-$. It is possible to evaluate the asymmetry as
\begin{eqnarray}
a_z=\frac{z^+-z^-}{z^++z^-}\approx\frac{1-|\frac{q_D}{p_D}|^2}{1+|\frac{q_D}{p_D}|^2},  \label{EQ:cpvxd}
\end{eqnarray}
where $z$ is $x_D$ or $y_D$. The same study may be applied assuming that $z$ is $y_{CP}$, $y'$, $x''$, $y''$.
The measured asymmetry would depend on the kind of $CP$ violation: if there is $CP$ violation in the decay, then the asymmetry would depend on the decay channel, while if $CP$ violation originates from mixing, then one would expect to obtain the same asymmetry in all modes. Estimates of the expected uncertainties that Super$B$ may obtain by combining different modes are given in Table~\ref{tbl:xdviolation}.
\begin{table}[!ht]
\caption{Combination of estimated uncertainties in the CPV mixing parameter $|q_D/p_D|$ that may be obtained at Super$B$ when considering the effective values of the mixing parameters and considering a 75 $ab^{-1}$ of data collected at the $\Upsilon(4S)$.}
\label{tbl:xdviolation}

\begin{center}
    \begin{tabular}{ | l | c  c  |}
    \hline
    Parameter 	&      	Mode		&	$\sigma(|q_D/p_D|\times10^2)$       \\ \hline \hline
    $x_D$	&	all modes	&	$\pm1.8$ \\
    $y_D$	&	all modes	&	$\pm1.1$ \\
		
    \hline
    \end{tabular}
\end{center}
\end{table}

\subsection{Direct $CP$ Violation}
The LHCb Collaboration has recently reported the first observation, at a $3.5 \sigma$ level, of time-integrated $CP$ asymmetry by combining the measurement for $D^0\to K^+K^-$ and $D^0\to \pi^+\pi^-$~\cite{note}. The result has been confirmed by the CDF Collaboration at a $2.7\sigma$ level.
In order to evaluate this asymmetry, one defines the asymmetry as shown in Eq.~(\ref{EQ:direct}): 

\begin{eqnarray}
A_{CP}(h^+h^-)=\frac{\Gamma(D^0 \to h^+h^-)-\Gamma(D^0 \to h^+h^-)}{\Gamma(D^0 \to h^+h^-)+\Gamma(D^0 \to h^+h^-)} &\approx& \nonumber \\ \approx A_{CP}^{dir}(H^+h^-)+\frac{<t(h^+h^-)>}{\tau}A_{CP}^{ind},  \label{EQ:direct}
\end{eqnarray} 
where the asymmetry is to first order a linear combination of the contribution coming from direct $CP$ violation ($A_{CP}^{dir}$), and from indirect $CP$ violation ($A_{CP}^{ind}$). The quantity $t(h^+h^-)$ is the mean of proper decay time of the $D^0$ meson decaying to the pair $h^+h^-$ and $\tau$ is the $D^0$ meson lifetime. One can compare the asymmetry in the two different channels $D^0\to K^+K^-$ and $D^0\to \pi^+\pi^-$ obtaining Eq.~(\ref{EQ:deltaa}):
\begin{eqnarray}
\Delta A_{CP}(h^+h^-)= A_{CP}(K^+K^-)-A_{CP}(\pi^+\pi^-)&=& \nonumber \\ = \Delta A_{CP}^{dir}+\frac{\Delta<t>}{\tau}A_{CP}^{ind}.  \label{EQ:deltaa}
\end{eqnarray} 
In Eq.~(\ref{EQ:deltaa}) the decay time difference is small, so that most of indirect $CP$ violation contributions cancel out, and then the expression for $\Delta A_{CP}$ approximates the difference in terms of direct $CP$ asymmetries between the two decays considered. Results of the analysis carried out by the LHCb and CDF Collaborations are given in Eqns.~(\ref{EQ:deltalhc})(\ref{EQ:deltacdf})~\cite{note}\cite{cdf}:
\begin{eqnarray}
\Delta A_{CP}(LHCb)= -0.82 \pm 0.21^{stat.}\pm 0.11^{sys.},  \label{EQ:deltalhc}
\end{eqnarray} 
\begin{eqnarray}
\Delta A_{CP}(CDF)= -0.62 \pm 0.21^{stat.}\pm 0.10^{sys.}.  \label{EQ:deltacdf}
\end{eqnarray} 

\begin{figure}[!ht]
\begin{center}
\resizebox{10.cm}{!}{
\includegraphics{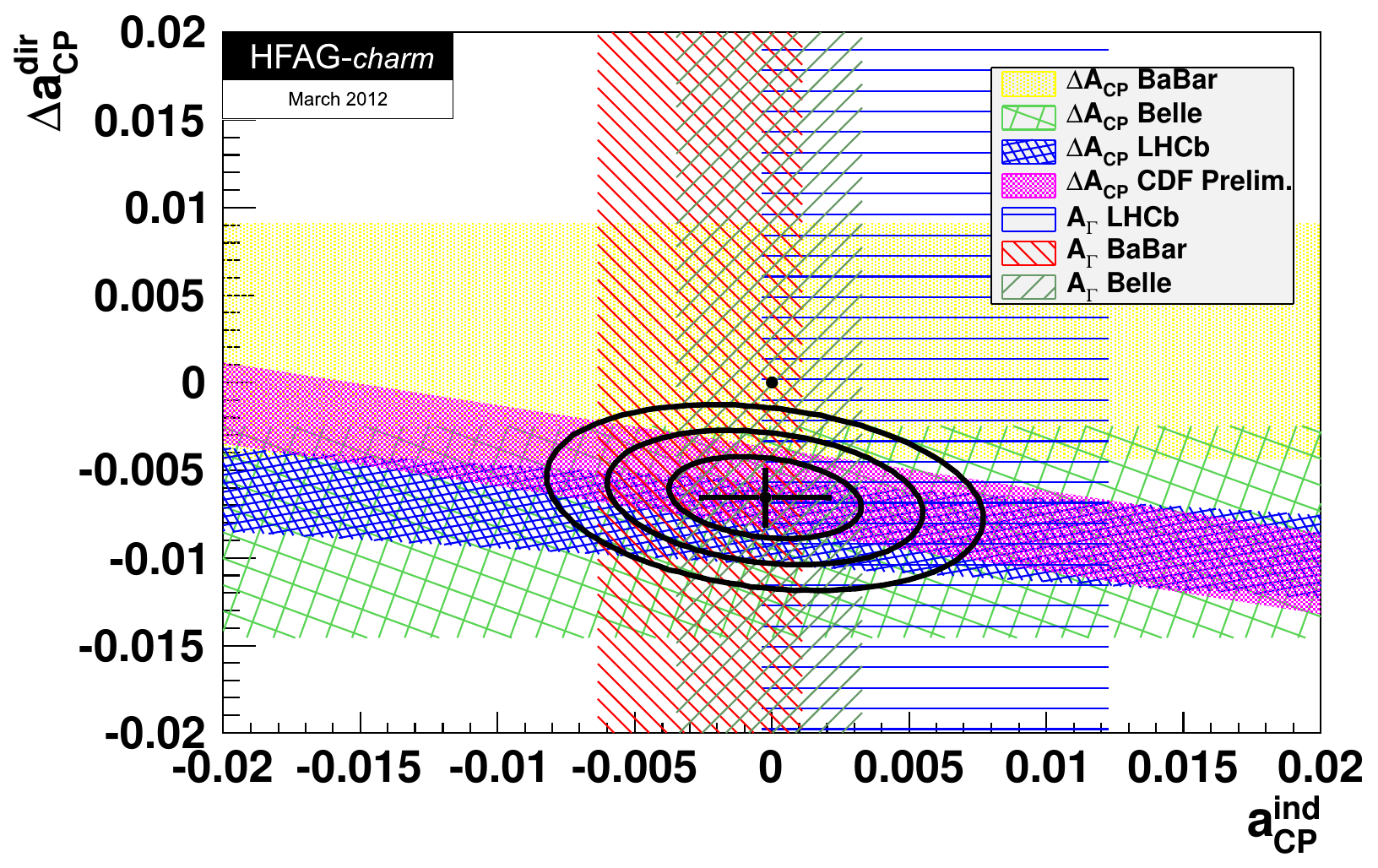}
}
\caption{HFAG combined results for $\Delta A_{CP}$.}\label{fig:hfag}
\end{center}
\end{figure}

The results from LHCb and CDF, together with the results coming from other experiments have been combined by the $heavy$ $flavour$ $averaging$ $group$ (HFAG) and are shown in Fig.~\ref{fig:hfag}, where data are consistent with the $CP$ conserving hypothesis at $0.006\%$ CL~\cite{hfag}.

The CDF collaboration has reported the results for the individual channels obtaining Eqns.~(\ref{EQ:cdf1})(\ref{EQ:cdf2}), where the naive SM expectation for the single channel is of the order of $10^{-5}-10^{-4}$
\begin{eqnarray}
A_{CP}(\pi^+\pi^-)= (+0.22 \pm 0.24^{stat.}\pm 0.11^{sys.})\%,  \label{EQ:cdf1}
\end{eqnarray} 
\begin{eqnarray}
A_{CP}(\pi^+\pi^-)= (-0.24 \pm 0.22^{stat.} \pm 0.09^{sys.})\%.  \label{EQ:cdf2}
\end{eqnarray} 
One should notice that the reported value for $\Delta A_{CP}$ is almost two orders of magnitude larger than naively predicted by the SM. Since it is not clear if the SM may account for such a big value of $\Delta A_{CP}$, new measurements (including also neutral final state particles) with better precision are needed in order to understand if this results is really to NP~\cite{kagan}.  
If one considers this approach to the study of $CP$ violation, the Super$B$ experiment can reach a sensitivity of the order of $\sigma=3\times10^{-4}$~\cite{superb1}.

\subsection{Time-dependent $CP$ Violation}
In Ref.~\cite{bim} time-dependent $CP$ violation (TD$CP$V) studies have been proposed for charm using a very similar formalism to that adopted when studying $B_d$ mesons. Since the Super$B$ experiment will collect data at the charm threshold (correlated $D^0$ mesons) and at the center-of-mass energy of the $\Upsilon(4S)$ (un-correlated $D^0$ mesons) it will then be possible to perform and combine time-dependent measurements obtained in the two different configurations of the machine. The method can be used also in an hadron machine (for example LHCb). Observations of TD$CP$V in charm can be used to constrain the angle $\beta_{c,eff}$ in the $charm$ $unitarity$ $triangle$, or $charm$ $triangle$, and the time-dependent analysis in general can be used to measure mixing parameters when TD$CP$V is observed or, when no $CP$ is observed it is possible to measure the mixing phase ($\phi_{MIX}$). Only the case of Super$B$ running at the $\Upsilon(4S)$ is discussed here. A more detailed description is available in Ref.~\cite{bim}.
\subsubsection{The Charm Unitarity triangle}
Unitarity of the CKM matrix allows one to write to six unitarity relations. One of the relations is obtained by combining the first two rows:
\begin{eqnarray}
V_{ud}^* V_{cd} + V_{us}^* V_{cs} + V_{ub}^* V_{cb} = 0, \label{eq:charmtriangle}  
\end{eqnarray}
where Eq.~(\ref{eq:charmtriangle}) defines the $charm$ $triangle$. The internal angles of this triangle are 
\begin{eqnarray}
\alpha_c &=& \arg\left[- V_{ub}^*V_{cb}/ V_{us}^*V_{cs} \right], \label{eq:alphac}\\
\beta_c  &=& \arg\left[- V_{ud}^*V_{cd} / V_{us}^*V_{cs} \right], \label{eq:betac}\\
\gamma_c &=& \arg\left[- V_{ub}^*V_{cb} / V_{ud}^*V_{cd}\right], \label{eq:gammac}
\end{eqnarray}
and using the results of Global CKM fits, one expects that:
\begin{eqnarray}
    \beta_c &=& (0.0350\pm 0.0001)^{\circ}. 
\end{eqnarray}
In Ref.~\cite{bim} the measurement of $\beta_{c,eff}$ using time-dependent $CP$ asymmetries in charm decays has been proposed by considering the decay channels $D^0 \to K^+K^-$ and $D^0 \to \pi^+\pi^-$. In fact neglecting LD contributions, the mentioned decay channels are tree dominated. In the decay $D^0 \to K^+K^-$ one would not expect to observe $CP$ violation, in particular this mode can be used to measure $\phi_{MIX}$. The situation is different for $D^0 \to \pi^+\pi^-$ in which not only a measurement of $\phi_{MIX}$ is possible, but since the decay is dominated by the CKM element $V_{cd}$ (which become complex when the CKM matrix is expanded up $\lambda^5+ O(\lambda^6)$) one would expect to measure $CP$ violation. Finally the accessible phase in $D^0 \to \pi^+\pi^-$ is $\phi_{MIX}-2\phi_{CP}$. If one subtracts the measured phase in $D^0 \to \pi^+\pi^-$ from the measured phase in $D^0 \to K^+K^-$ then a measurement of $\beta_{c,eff}$ is obtained. 
\subsection{Time-dependent formalism}
As anticipated earlier, only un-correlated $D^0$ mesons are considered. Un-correlated $D^0$'s are produced from the decays of $B$ mesons in electron-positron colliders when particles are collided at a center of mass energy corresponding to the $\Upsilon(4S)$ resonance, from $c\overline{c}$ continuum, or in hadrons colliders. 

The time-dependent asymmetry for un-correlated charm mesons can be expressed by 

\begin{eqnarray}
{\cal A}^{Phys}(t) = \frac{\overline{\Gamma}^{Phys}(t) - \Gamma^{Phys}(t) } { \overline{\Gamma}^{Phys}(t) + \Gamma^{Phys}(t)} &=& \nonumber \\= -\Delta \omega + \frac{ (D + \Delta\omega)e^{\Delta \Gamma t/2}[ (|\lambda_{f}|^2 - 1)\cos\Delta M t + 2 Im\lambda_{f} \sin\Delta M t ]}{h_+ (1+|\lambda_{f}|)^2/2 + Re(\lambda_{f}) h_-}.
  \label{eq:asymtagging1}
\end{eqnarray}

Where $\Delta \Gamma$ and $\Delta M$ are the width and mass differences between the $D_H$ and $D_L$ strong eigenstates (H and L indicates the heavy and the light eigenstate), $\omega$ and $D$ are the mistag probability and the dilution, respectively, $\lambda_f= \frac{q}{p} e^{i\phi_{MIX}} \frac{\bar A}{A} e^{i\phi_{CP}}$.
The charm mixing parameters are defined by
\begin{eqnarray}
x_D=\frac{\Delta M}{\Gamma}, y_D= \frac{\Delta \Gamma}{2\Gamma}.  \label{EQ:xandy}
\end{eqnarray}
The definition of the charm mixing parameters and the form of the time-dependent asymmetry allows one to test $CP$ violation in terms of $x_D$ and $y_D$ providing a measurements of mixing.
The time-dependent asymmetry expressed in terms of $x_D$ and $y_D$ becomes~\cite{myself}:

\begin{eqnarray}
{\cal A}^{Phys}_{x_D,y_D}(t) = -\Delta \omega + \frac{ (D + \Delta\omega)e^{y_D\Gamma t}[ (|\lambda_{f}|^2 - 1)\cos x_D\Gamma t + 2 Im\lambda_{f} \sin x_D\Gamma t ]}{h_+ (1+|\lambda_{f}|)^2/2 + Re(\lambda_{f}) h_-}.
  \label{eq:asymtaggingxy}
\end{eqnarray}

\subsection{Sensitivity to \boldmath{$\beta_{c,eff}$}, \boldmath{$\phi_{MIX}$}, \boldmath{$\phi_{CP}$ and $x_D$} }
Expected sensitivities on the parameters described in the previous section are reported here for un-correlated $D^0$ mesons production. More detailed results (including other experiments) are given in Ref.~\cite{bim}. In Table~\ref{tbl:numericalanalysis} the expected uncertainties for $\phi_{MIX}$ and $\sigma_{\beta_{c,eff}}$ are shown, and Table~\ref{tbl:deltax} shows the expected sensitivities on $x_D$.
\begin{table}[!ht]
\caption{Summary of expected uncertainties from 75 $ab^{-1}$ of 
data at the $\Upsilon(4S)$ for $D^0\to\pi^+\pi^-$ decays.}
\label{tbl:numericalanalysis}

\begin{center}
    \begin{tabular}{ | l | c |}
    \hline
    {Parameter} &                  Un-correlated $D$'s ($\Upsilon(4S)$) \\ \hline\hline
    $\sigma_{\phi_{\pi\pi}}=\sigma_{arg(\lambda_{\pi\pi})}$ &  $2.2^\circ$  \\  
    $\sigma_{\phi_{KK}}=\sigma_{arg(\lambda_{KK})}=\sigma{\phi_{MIX}}$ &  $1.6^\circ$ \\ 
    $\sigma_{\beta_{c,eff}}$ &  $1.4^\circ$ \\
    \hline
    \end{tabular}
\end{center}
\end{table}
\begin{table}[!ht]
\caption{Estimates of the sensitivity on $x_D$ when a $75 ab^{-1}$ data collected at the $\Upsilon(4S)$ is considered for the decays $D^0 \to \pi^+ \pi^-$ and $D^0 \to K^+ K^-$ and $\phi=\phi_{MIX}-2\beta_{c,eff}$.}\label{tbl:deltax}
\begin{center}
    \begin{tabular}{ | l | c | c |}
    \hline
    Mode/HFAG & $\sigma_{x_D} (\phi = \pm 10^o)$ & $\sigma_{x_D} (\phi = \pm 20^o)$  \\ \hline \hline
    Super$B$ [$\Upsilon(4S)$]  &  & \\ 
    $D^0 \to \pi^+ \pi^-$ & $0.12\%$& $0.06\%$      \\  
    $D^0 \to   K^+   K^-$ & $0.08\%$& $0.04\%$      \\ \hline \hline
    HFAG & \multicolumn{2}{|c|}{$0.20\%$} \\
  \hline
    \end{tabular}
\end{center}
\end{table}

This shows that a time-dependent analysis applied to charm not only may help us to better understand the flavour changing structure of the SM, but the observation of $CP$ violation in charm may allow us to study the charm triangle through the measurement of the angle $\beta_{c,eff}$. This will provide us with a consistency check of the CKM mechanism. It is important to mention that with current experimental sensitivity, any observation of a value of the $\beta_{c,eff}$ inconsistent with zero will be a clear signal of new physics. 
On the other hand, the time-dependent analysis can measure $x_D$ and $\phi_{MIX}$ using not only the decay modes $D^0\to K^+K^-$ or $D^0 \to \pi^+\pi^-$, but additional decay modes are available to carry out time-dependent mixing-related measurements. Decay channels including neutral final states, as it is the case of $D^0 \to K_S^0 \pi^0$, with branching ratios that are larger than those for $D^0\to K^+K^-$ and $D^0\to \pi^+\pi^-$, may provide better constraints in the determination of mixing phase/parameters, and this makes an electron-positron machine unique when performing such a measurement.

\section{Conclusions}
\label{sec:conclusions}
In this work, the potential reach of the Super$B$ experiment when studying charm physics has been discussed. Rare decays have been presented from a phenomenological point of view and it was shown that they can be used not only when looking for new physics but also when trying to interpret new physics signals. Charm mixing has been introduced and it was shown how the Super$B$ experiment, will be able to improve the precision on $x_D$ and $y_D$, this is not only due to the large luminosity that is expected to be collected at $\Upsilon(4S)$. The Super$B$ experiment with its run at the charm threshold will be able to perform strong phase measurements that will shrink the allowed parameter space, and it has been shown that by using the $effective$ $values$ for the mixing parameters it is possible to perform a test of $CP$ violation (indirect). Direct $CP$ violation has been discussed, and it has been highlighted that it is not yet clear if the standard model may account for the observed asymmetry, or if some new physics is showing up, and studies of additional modes help resolve this question. Super$B$ may play a leading role in the understanding of direct $CP$ violation not only repeating the analysis already carried out, but performing new analyses that would otherwise be difficult for other experiments (final states including neutrals). Finally, time-dependent $CP$ violation has been introduced and it was shown that with that formalism it is possible to perform measurements of the quantities: $\phi_{MIX}$, $x_D$, and $\beta_{c,eff}$. The latter not only represents a missing piece in the study of the CKM mechanism, but it can be used to search for new physics. It was shown that with current sensitivities (running and planned experiments), any observation of of a value of $\beta_{c,eff}$ different from zero, will be a clear signal of new physics.

\section{Acknowledgments}
This work has been supported by Queen Mary, University of London Postgraduate Research Fund. I wish to thanks the organisers of this workshop and the University of Hawai'i for their warm welcome and for local support.

\end{document}